\documentclass[a4paper,11pt]{article}
 \providecommand{\tightlist}{%
     \setlength{\itemsep}{0pt}\setlength{\parskip}{0pt}}
\usepackage{graphicx}
\usepackage{jheppub} 
\usepackage{comment} 
\usepackage[dvipsnames]{xcolor}
\usepackage{tikz}
\usepackage{slashed}
\usepackage{pgfplots}
\pgfplotsset{compat=1.14}
\usepackage{booktabs, longtable}
\usepackage{microtype}
\usepackage{soul}
\renewcommand{\H}{\widetilde{H}^0}
\newcommand{\W}{\widetilde{W}^0}
\newcommand{\B}{\widetilde{B}^0}
\newcommand{\N}{\widetilde{\chi}^0}
\newcommand{\C}{\widetilde{\chi}^\pm}

\newcommand{\met}{{\slashed{E}_T}}
\newcommand{\cb}{ c_\beta}
\newcommand{\cw}{ c_{\theta_W}}
\newcommand{\sinb}{ s_\beta}
\newcommand{\sw}{ s_W}
\newcommand{\mz}{ m_Z}

\title{Higgs Assisted Razor Search for Higgsinos at a 100 TeV $pp$ Collider}
\author[a]{Adarsh Pyarelal}
\author[b]{and Shufang Su}
\affiliation[a]{School of Information, University of Arizona, Tucson, AZ 85721 , USA}
\affiliation[b]{Department of Physics, University of Arizona, Tucson, AZ 85718, USA}

\emailAdd{adarsh@email.arizona.edu}
\emailAdd{shufang@email.arizona.edu}

\abstract{%
  A 100 TeV proton-proton collider will be an extremely effective way to probe
  the electroweak sector of the Minimal Supersymmetric Standard Model (MSSM).
  In this paper, we describe a search strategy for discovering pair-produced
  Higgsino-like next-to-lightest supersymmetric particles (NLSPs) at a 100 TeV
  hadron collider that decay to Bino-like lightest supersymmetric particle
  (LSP) via intermediate $Z$ and SM Higgs boson that in turn decay to a pair of
  leptons and a pair of $b$-quarks respectively: $\N_{2}\N_{3} \rightarrow
  (Z\N_1)(h\N_1)\rightarrow bb\ell\ell+\N_1\N_1$. In addition, we examine the
  potential for machine learning techniques to boost the power of our searches.
  Using this analysis, Higgsinos up to 1.4 TeV can be discovered  at 5 $\sigma$
  level for a Bino with mass of about 0.9 TeV using 3000 fb$^{-1}$ of data.
  Additionally, Higgsinos up to 1.8 TeV can be excluded  at 95\% C.L. for Binos
  with mass of about 1.4 TeV. This search channel extends the multi-lepton
  search limits, especially in the region where the mass difference between the
  Higgsino NLSPs and the Bino LSP is small.
}

\begin{document}

\maketitle

\section{Introduction}

The existence of dark matter provides unambiguous evidence of new physics
beyond the Standard Model (SM) of particle physics. Indeed, given that there is
far more dark matter in the Universe than there is baryonic matter, determining
its precise nature is one of the most exciting challenges in particle physics
today.  One promising group of candidates for dark matter is the class of stable
particles known as weakly interacting massive particles (WIMPs). These
particles arise naturally in many extensions of the SM that are motivated by
the need to stabilize the electroweak scale. One of the most heavily studied extensions
is the Minimal Supersymmetric Standard Model (MSSM), which predicts a
`supersymmetric partner', or \emph{superpartner}, for each SM
particle\footnote{For a detailed review of the MSSM, see Ref.
\cite{Martin:1997ns}.}. In the MSSM with \emph{R}-parity conservation, the
lightest supersymmetric partner (LSP) is predicted to be absolutely stable,
making it a good candidate for dark matter. The identity of the LSP is
determined by the mass hierarchy of the superpartners, which in turn depends on
the mechanism of supersymmetry breaking.  However, taking into account experimental
constraints and phenomenological considerations, the lightest of the MSSM
particles known as the neutralinos emerges as the most attractive candidate
for the LSP \cite{Bertone:2004pz}. 

Though a natural MSSM spectrum has the potential to tame the hierarchy problem,
it is under siege from recent data from the Large Hadron Collider (LHC)
\cite{Aaboud:2018ujj, Sirunyan:2018vjp}.  A compelling alternative scenario
comes in the form of split supersymmetry (split SUSY) \cite{Wells:2003tf,
ArkaniHamed:2004yi, Giudice:2004tc}. In this scenario, the lightest
superpartners are the fermionic ones, on the scale of 1-10 TeV, while the
scalar superpartners can be much heavier, on the scale of 100 - 1000 TeV. In
exchange for accepting some level of fine-tuning, we obtain numerous benefits,
including the suppression of flavor-changing neutral currents and greater
compatibility with data from CP-violation experiments.  The low lying fermionic
superpartners consist of neutralinos and charginos (collectively known as
electroweakinos), and gluinos. The current LHC search limits on gluinos are
already around 2 TeV for prompt decay and 2.5 TeV for disappearing track
searches~\cite{CMS-PAS-SUS-19-005}.   The limits on electroweakinos with heavy
scalar superparticles, however, are still relatively weak: around 650 $-$ 750
GeV or less~\cite{ATL-PHYS-PUB-2019-022, Sirunyan:2017lae, Aaboud:2018htj,
Sirunyan:2018ubx, Aaboud:2018zeb}, depending on the decay modes of the parent
electroweakinos and their species.  In this paper, we will focus on the
discovery potential of the neutralinos and charginos at a 100 TeV $pp$
collider. 

The decay patterns of the electroweakinos (Winos, Bino, and Higgsinos) depend
on the hierarchy of the MSSM parameters $M_1, M_2$, and $\mu$~\footnote{For our
study, we assume all the mass parameters are positive.  For the current
analyses, the results do not depend much on the sign of those parameters.},
which roughly determine the masses of the Bino, Winos, and Higgsinos,
respectively~\cite{Han:2013kza}.  In this paper, we consider the Higgsino NLSPs
and Bino LSP scenario with $M_1 < \mu \ll M_2$.  One of the motivations for
studying Higgsinos is that the parameter $\mu$ that governs their masses plays an
important role in electroweak symmetry breaking~\cite{Acharya:2014pua}, and
is driven to be relatively small by naturalness considerations. 
Finding Higgsinos has traditionally been more challenging than finding Winos,
due to their smaller production rates.
This, however, can be remedied by the increased production cross sections and 
luminosities at a 100 TeV $pp$ collider.

Since the Bino is the only supersymmetric particle lighter than the Higgsinos in
the mass hierarchy we consider, the neutral Higgsinos will decay to a Bino and a
neutral SM boson (\emph{Z/h}) with a branching fraction of 100\%.  Searches for
Higgsino NLSPs with Bino LSPs via  multi-lepton final states have been studied
in Ref.~\cite{Gori:2014oua} for a 100 TeV $pp$ collider, for pair production of
$\N_{2,3}\C_1$ and
$\widetilde\chi_{1}^+\widetilde\chi_1^-$. In this paper, we attempt to exploit
the relatively large branching fraction of the SM Higgs boson to \emph{b} quarks,
and study the potential of the $\N_2\N_3\rightarrow Zh
\N_1\N_1\rightarrow bb\ell\ell \met$ channel.

A 100 TeV $pp$ collider would open up an immense number of physics
opportunities not afforded to the 14 TeV LHC, including exploring the most
interesting regions of the split SUSY parameter space
\cite{Arkani-Hamed:2015vfh}.
Such a collider would be a natural next step after the LHC, and is being
actively discussed in the particle physics community, with two major proposals
being the Future Circular Collider (FCC)-hh by CERN \cite{FCC-hh}, and the
Super Proton Proton Collider (SppC)  in China \cite{CEPC}.  As the next energy
frontier machine, it is crucial to fully explore the physics
potential of a 100 TeV $pp$ collider, especially for new particles that are
either too heavy or too rare to be produced at the
LHC~\cite{Arkani-Hamed:2015vfh,Contino:2016spe,Golling:2016gvc,Mangano:2016jyj}.
In particular, studies have already been performed on the prospects of
discovering neutralino dark matter for compressed \cite{Low:2014cba,
diCortona:2014yua,Cirelli:2014dsa,Mahbubani:2018tin,Han:2018wus} and
well-separated \cite{Gori:2014oua,Acharya:2014pua} neutralino spectra.

In our study, we use razor variables~\cite{Rogan:2010kb}, which were originally designed
for searches involving two heavy, mass-degenerate pair-produced
particles, each of which decays into a visible and invisible set of
particles. This topology matches that of our search channel, making this
set of variables a natural choice for our analysis.  

At a 100 TeV collider, the  SM backgrounds are going to be even larger than
at the LHC.  An ancillary goal of this paper is to investigate the
potential for machine learning (ML) techniques to augment our analysis.
With the advent of more powerful computers and simultaneous advances in
the field of statistical learning in recent years, the usage of ML is
rising in experimental particle physics -- in fact, the discovery of the
SM Higgs boson in 2012 was done with the help of neural networks
\cite{Aad:2012tfa} and boosted decision trees (BDTs)
\cite{Chatrchyan:2012xdj}.

The rest of the paper is structured as follows. In Sec.~\ref{sec:model}, we
describe our model and search channel in more detail, and review the existing
experimental search bounds. In Sec.~\ref{sec:analysis}, we describe our
analysis strategies for both traditional cut-and-count analysis and analysis
performed using BDTs. In Sec.~\ref{sec:results}, we present the results of our
analyses, i.e. the expected 5$\sigma$ discovery and 95\% C.L. exclusion
reaches. Finally, we conclude with the implications in
Sec.~\ref{sec:conclusion}.

\section{The neutralino sector of the MSSM}
\label{sec:model}

The neutralino sector of MSSM consists of four mass eigenstates
($\N_1,\N_2,\N_3,\N_4$), which are mixtures of the Bino, Wino, and two neutral
Higgsinos: ($\B,\W,\H_d,\H_u$). In the basis of these gauge
eigenstates, the mass matrix of the neutralinos can be written as
\begin{align}
  \mathbf{M}_{\widetilde{N}}=
  \begin{pmatrix}
    M_1         & 0            & -\cb\sw\mz & \sinb\sw\mz \\
    0           & M_2          & \cb\cw\mz  & -\sinb\cw\mz \\
    -\cb\sw\mz  & \cb\cw\mz    & 0          & -\mu \\
    \sinb\sw\mz & -\sinb\cw\mz & -\mu       & 0
  \end{pmatrix},
\end{align} 
where we employ the notation $s_\theta, c_\theta = \sin\theta, \cos\theta$. Here,  $\theta_W$ is the Weinberg mixing angle, 
and $\tan\beta= v_u/v_d$, where $v_u$ and $v_d$ are the vacuum
expectation values of $H_u^0$ and $H_d^0$, with $v_u^2+v_d^2=v^2=(246\ {\rm
GeV})^2$. The parameters $M_1$, $M_2$, and $\mu$ are the mass parameters of
the Bino, Wino, and the Higgsinos, respectively.
The mass eigenstates are labeled as $\N_i$ for $i=1 \ldots 4$ with increasing mass eigenvalues.
In the limit where $\mz \ll |\mu\pm M_1|, |\mu\pm M_2|,$ the mass eigenstates
are, to good approximation, a nearly pure Bino, $\B$, with mass $M_1$,
a nearly pure Wino, $\W$, with mass $M_2$, and nearly pure Higgsinos
$\H_{1,2} = (\H_u \pm \H_d)/\sqrt{2}$, with mass
$|\mu|$. 

The optimal search strategy for finding electroweakinos   highly depends on
their mass spectra \cite{Han:2013kza}.  Phenomenological collider studies on
finding electroweakinos with a small mass splitting of about 0.1 $-$ 50 GeV at a
100 TeV $pp$  collider can be found in \cite{Low:2014cba, Bramante:2014tba,
Berlin:2015aba, Cirelli:2014dsa}.
For $3\ \text{ab}^{-1}$ of data, the projected exclusion reaches for pure Wino
LSPs using the monojet channel is about $1.4$~TeV.  The reach
can be increased to $\sim3$ TeV with a disappearing tracks search.
Neutralinos up to 1 TeV can be excluded for mass splittings of about 20-30 GeV,
using a soft lepton search \cite{Low:2014cba}. Additionally, Bino-Wino LSPs
with masses up to 1.5 TeV for an inter-state mass splitting of around 1 GeV  can be
discovered at a 100 TeV $pp$ collider with $\sim7\ \text{ab}^{-1}$
of data \cite{Bramante:2014tba}.

Well-separated spectra have been studied in \cite{Gori:2014oua,
Acharya:2014pua} using multi-lepton channels, of the form 

\begin{align}
  pp\rightarrow \widetilde{\chi}_i\widetilde{\chi}_j\rightarrow VV +
  \N_1\N_1\rightarrow 2\ell/3\ell + \met,
\end{align} 

\noindent where $\widetilde{\chi}_{i,j}$ can be a neutralino or chargino, $\N_1$
is the lightest neutralino, and $V$ is any of the SM gauge bosons $W$, $Z$
or scalar $h$.  These searches work well since the large mass difference
between the electroweakinos can lead to energetic leptons that can be easily
identified. Among the multi-lepton searches for Higgsinos, the trilepton
searches: $\N_{2,3}\C_1\rightarrow WZ\N_1\N_1\rightarrow \ell\ell\ell +
\met$, with $\ell=e, \ \mu$,  have the best reach, due to the relatively high
production cross-section of chargino-neutralino pairs combined with the large
reduction in $t\overline{t}$ and QCD backgrounds from requiring three
leptons.  Higgsino NLSPs can be excluded at 95\% C.L. with mass up to 2.8 TeV
or be discovered with  mass up to 1.8 TeV, for a massless Bino LSP.   The reach
reduces for larger $M_1$, and vanishes for $M_1\gtrsim 600$ GeV (1000 GeV)
for discovery (exclusion).  Multilepton searches with \emph{Zh} and \emph{ZZ}
as the intermediate dibosons are unlikely to be as powerful, due to the smaller
pair-production cross-section of neutral Higgsinos, combined with the low
branching fraction of \emph{Z/h} to leptons. However, if we move beyond
multilepton searches, the channel with \emph{Zh} (originally suggested in
Ref.~\cite{Han:2013kza}) as the intermediate dibosons emerges as a possible
competitor to the \emph{WZ} channel.  

\begin{figure}[h]
  \centering
  \includegraphics[width=0.49\textwidth]{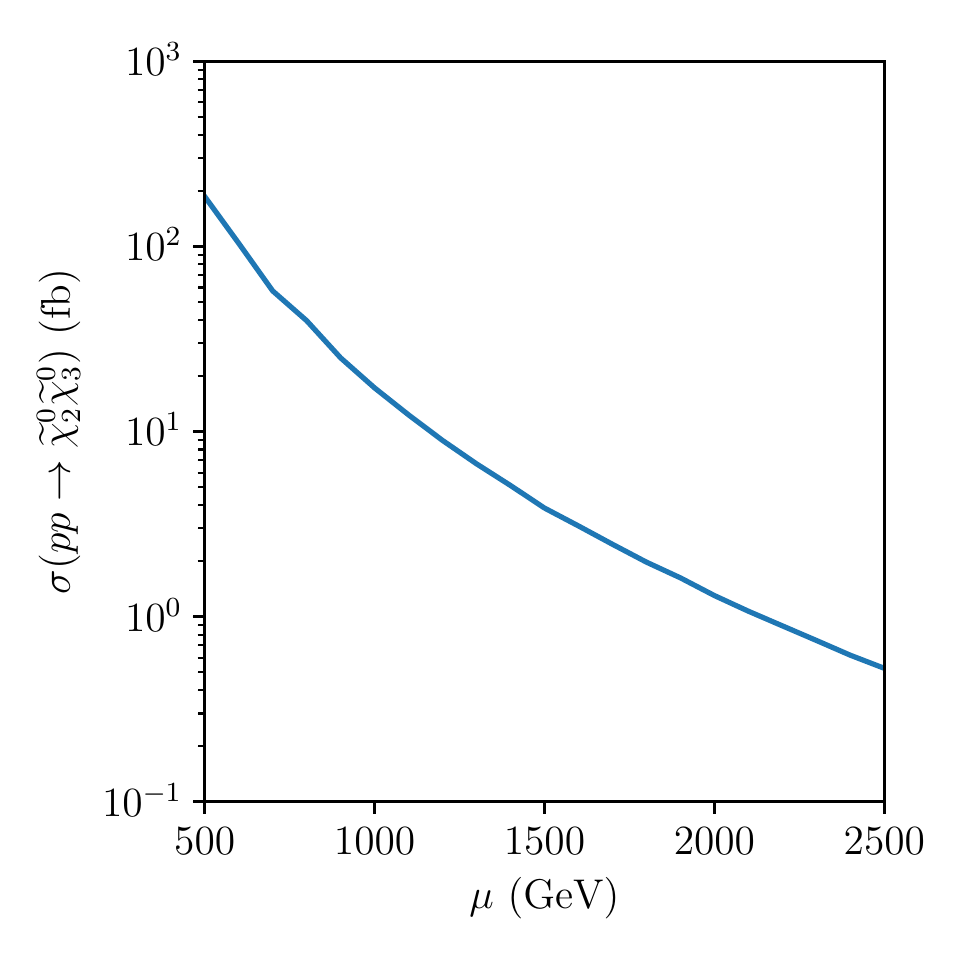}
  \includegraphics[width=0.49\textwidth]{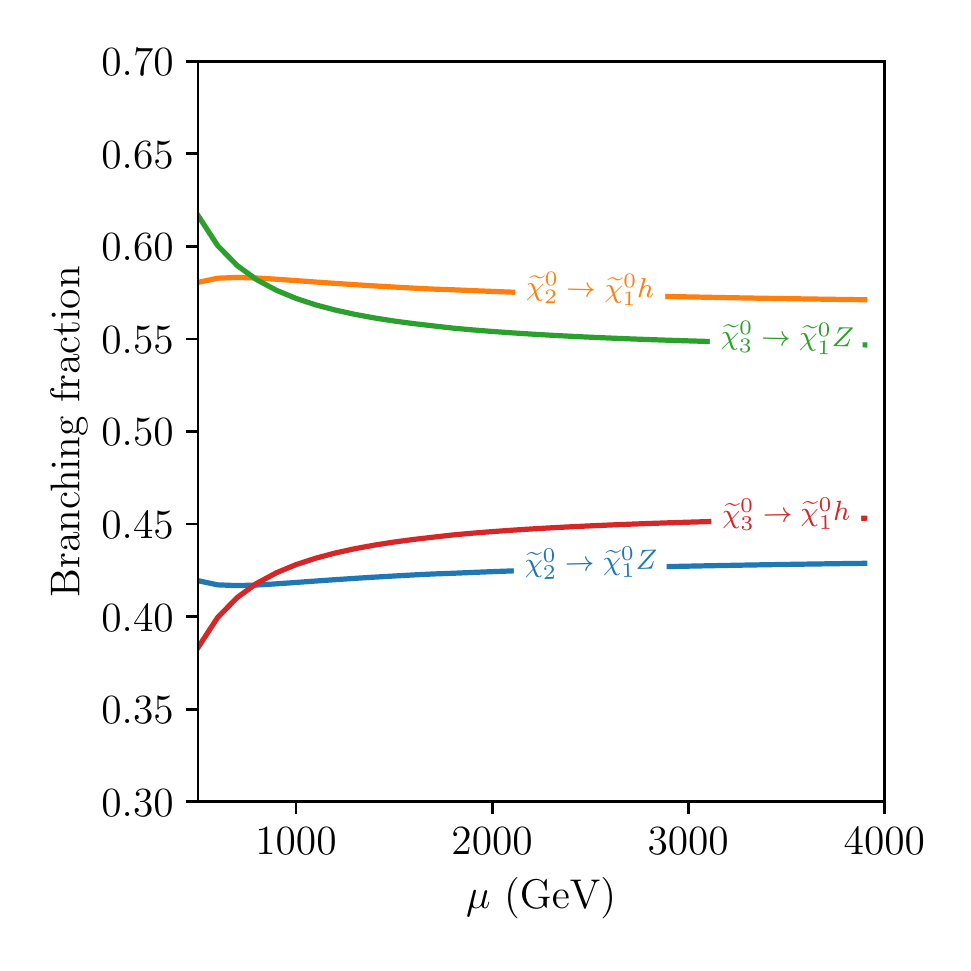}
  \caption{Higgsino pair production $\N_2\N_3$
    cross section (left panel), calculated using
    Prospino2~\cite{Beenakker:1999xh},  and decay branching fractions
    (right panel), calculated in SUSY-HIT~\cite{Djouadi:2006bz},  as a function
    of $\mu$ for $M_1=25$ GeV.
}
  \label{fig:xsection_plot}
\end{figure}

In the left panel of Fig.~\ref{fig:xsection_plot}, we show the pair-production
cross section $\N_2\N_3$ for the Higgsino-like NLSPs.  The cross section varies
between 200 fb to 0.5 fb for Higgsino masses between 500 GeV and 2500 GeV. The
right panel of Fig.~\ref{fig:xsection_plot} shows the decay branching fractions
of $\chi_{2,3}^0\rightarrow Z/h \chi_1^0$, as calculated by SUSY-HIT
\cite{Djouadi:2006bz}.  While one of the Higgsinos mostly decays to $Z$, the
other Higgsino mostly decays to $h$.    Therefore,  for pair produced Higgsinos
$\N_2\N_3$, the \emph{Zh} mode has the greatest branching fraction.  After
multiplying all the branching fractions, the signal cross sections of 

\begin{align}
  pp\rightarrow \N_2\N_3\rightarrow (Z\N_1)(h\N_1)\rightarrow bb\ell\ell \N_1\N_1.
\end{align} 

\noindent become comparable to the trilepton processes, as shown in Table
\ref{tab:xsections}. Thus, the $Zh+\met$ signal provides an alternative discovery
channel for Higgsinos. The signal contains two \emph{b}-jets, two same flavor,
opposite sign leptons, and missing transverse energy. The main backgrounds for
this process are: $t\overline{t}$, \emph{tbW} with the $b$ and $W$ not coming
from a \emph{t}, and \emph{bbWW} with no intermediate $s$-channel top quarks.

\begin{table}
  \centering
  \begin{tabular}{l|rr}
    \toprule
    Stage & $\N_{2,3}\C_1$ & $\N_2\N_3$\\
    \midrule
    Pair production cross section & 60 fb & 17.2 fb\\
    Intermediate diboson contribution & ($WZ$) 29.7 fb  & ($Zh$) 9.15 fb \\
    Applying \emph{BR}$(W\rightarrow \ell\nu)$, \emph{BR}$(Z\rightarrow \ell\ell)$ \& \emph{BR}$(h\rightarrow bb)$ & 0.42 fb & 0.37 fb\\
    \bottomrule
  \end{tabular}
  \caption{Comparison of Higgsino pair production cross-sections for
    $(\N_{2,3}\C_1)$ and
    $(\N_{2}\N_{3})$, for $\mu\approx 1$ TeV, at a 100
    TeV \emph{pp} collider. 
    The pair-production cross section for $\N_{2,3}\C_1$ is taken from
    \cite{Gori:2014oua}, while the pair-production cross-section for $\N_2\N_3$
    is calculated at Next-to-Leading Order (NLO) using Prospino2. The branching
    ratios to the diboson intermediate states are calculated using SUSY-HIT,
    and the branching fractions to the SM final states are taken from
    \cite{Olive:2016xmw}.}
  \label{tab:xsections}
\end{table}

Searches for pair produced nearly mass-degenerate Higgsinos  have been
performed at ATLAS and CMS for various decay topologies
\cite{Aaboud:2018htj,Sirunyan:2018ubx,Aaboud:2018zeb}.  
For neutralino-chargino pair production with \emph{WZ} and \emph{Wh} topologies, the 95\%
C.L. exclusion limits are about 650 GeV, 490 GeV, and 535 GeV for \emph{WZ}, \emph{Wh} and an equal mixture of \emph{WZ}+\emph{Wh} topology, with the strongest limits coming from
the multi-lepton searches.  For neutralino pair production with \emph{ZZ, Zh}, and \emph{hh} topologies\footnote{While such searches assume a gravitino LSP with Higgsino NLSPs, the limits are still relevant for our Higgsino NLSPs - Bino
LSP scenario.}, the limits are about 650 $-$ 750 GeV via multi-lepton or
multi-$b$ jets searches, depending on the decay branching fractions of the Higgsino NLSPs.


\section{Analysis Details}\label{sec:analysis}

In this section, we  describe our strategies for both the traditional
cut-and-count analysis, as well as the analysis carried out with boosted
decision trees. 

\subsection{Simulation}\label{simulation}

We simulated parton-level events using \texttt{MadGraph5 v2.3.2.2} and
\texttt{MadEvent} \cite{Alwall:2014hca}, then passed those events to Pythia 6
\cite{Sjostrand:2006za} for showering and hadronization. Finally, we used
\texttt{Delphes 3} \cite{deFavereau:2013fsa} to perform a fast, parametrized
detector simulation, with the detector card devised by the FCC-hh working
group\footnote{\url{https://github.com/HEP-FCC/FCCSW/blob/master/Sim/SimDelphesInterface/data/FCChh_DelphesCard_Baseline_v01.tcl}}.
For the backgrounds, we allowed up to one additional jet in the final state, to
approximate NLO QCD effects, and performed MLM matching with the \texttt{xqcut}
parameter set to 40 GeV. The Higgsino pair production cross sections were
calculated using \texttt{Prospino2}~\cite{Beenakker:1999xh} at NLO.  
To decouple the Wino, we set its mass parameter $M_2$ to 3 TeV.  

Since we expect our signal process to have a dilepton resonance from an on-shell
$Z$ boson, we restricted the phase space for event generation for backgrounds to
the region where the invariant mass of dilepton pairs lies between 80 and 100
GeV. Additionally, the Bino dark matter that escapes the detector would result
in a large amount of missing transverse energy ($\met$), so we required
a minimum $\met$ of 100 GeV at the parton level for the backgrounds as
well.  

At the detector simulation level, we relaxed the lepton isolation criterion in the
\texttt{Delphes} detector card from $\Delta R_{min}$ from 0.4 to 0.05. This is
motivated by the fact that due to the large mass difference between the Higgsino
NLSPs and the Bino LSP in our search channel, the intermediate $Z$ bosons will be
highly boosted, and the leptons to which they decay will be highly collimated.
The value of 0.05 is consistent with what is suggested in previous 100 TeV
studies~\cite{Acharya:2014pua,Gori:2014oua,Bramante:2014tba} and will allow
for easier comparison between different search strategies.

\subsection{Analysis using cut-and-count}
\label{event-selection}

For the cut-and-count analysis, we implemented successive one-dimensional cuts
on the variables listed below, using the MadAnalysis 5 package
\cite{Conte:2012fm}.

\begin{figure}[h]
\centering
\includegraphics[trim = {0.3cm 0.6cm 0.1cm 0}, clip, width=0.496\textwidth]{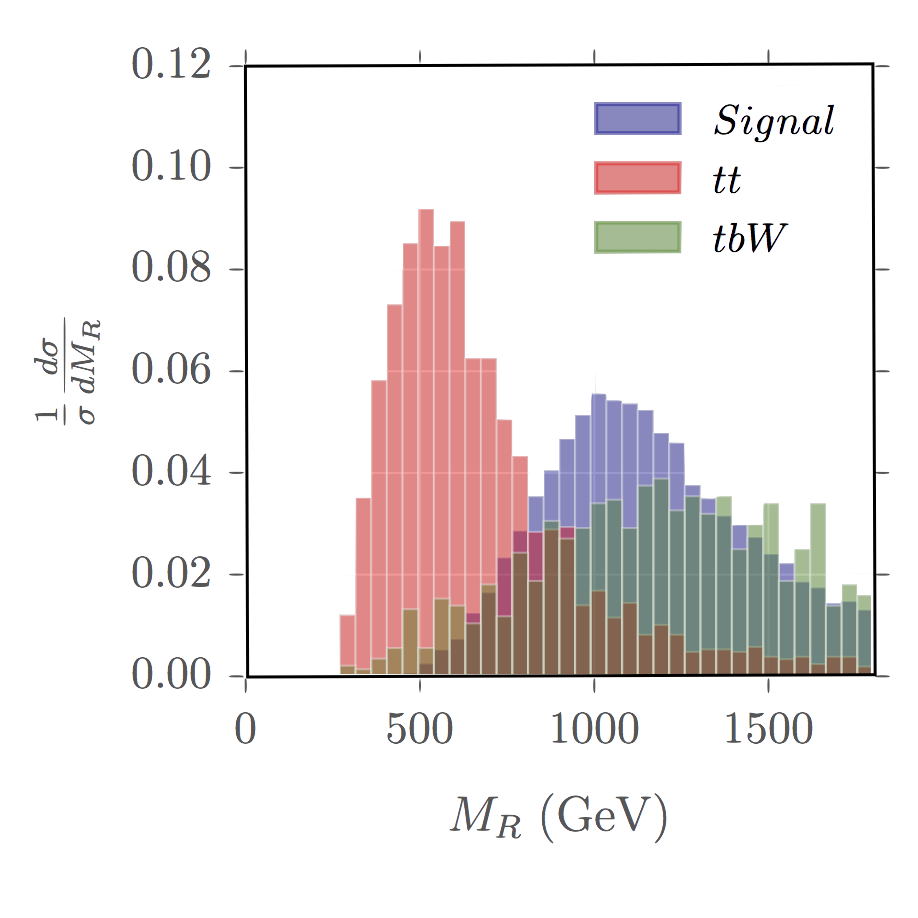}
\includegraphics[trim = {0.3cm 0.6cm 0.1cm 0}, clip, width=0.496\textwidth]{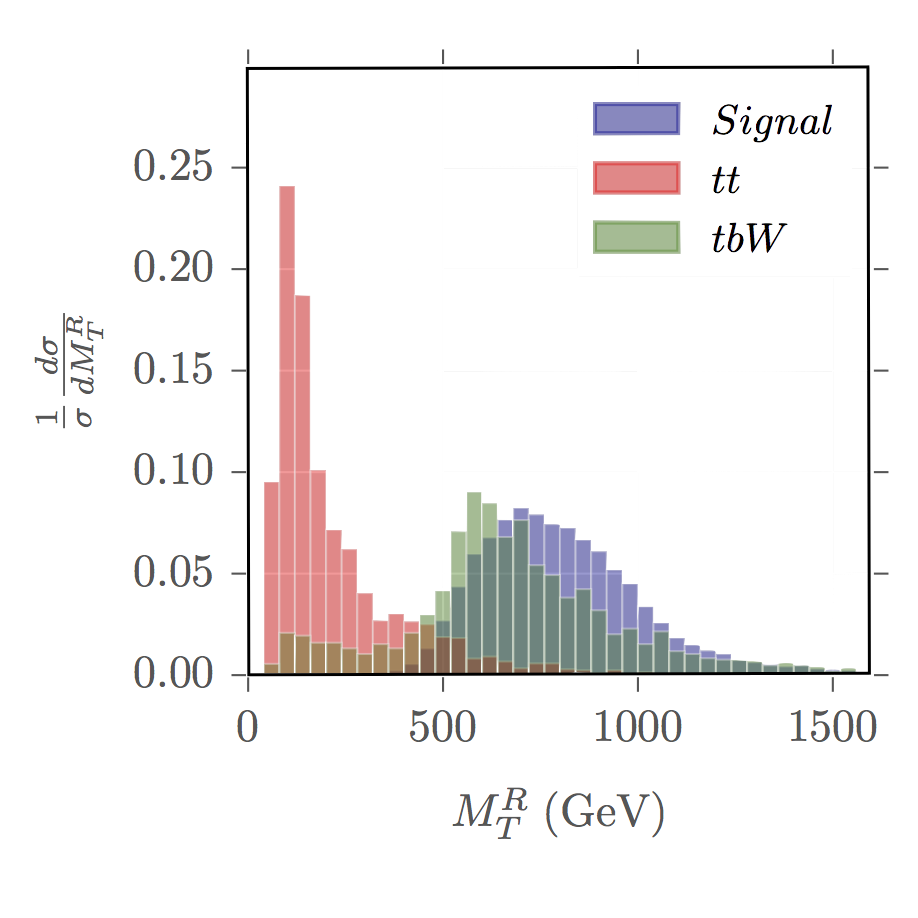}
\caption{Normalized distributions of the razor kinematic variables $M_R$
  (left) and $M_T^R$ (right) for  a signal benchmark point with $\mu=1$ TeV and $M_1=25$ GeV,
  as well as the dominant $tt$ and $tbW$ backgrounds
  after the trigger and identification cuts.}
\label{fig:razor_histos}
\end{figure}

\begin{enumerate}
  \tightlist
  \item \emph{Trigger}: Events were selected if they had at least one lepton
    with $p_{T}^\ell >$ 100 GeV. 

  \item \emph{Identification:}

    \begin{itemize}
      \item Events contain exactly two leptons of the same flavor
        and with opposite signs (\textsc{sfos}), with $p_{T}^\ell >$ 15 GeV and
        $\vert\eta^\ell\vert <$ 2.5.
      \item Events contain at least two \emph{b}-tagged jets with
        $p_{T}^b > 30$ GeV and $\vert\eta^b\vert <$ 2.5.
    \end{itemize}

   \item \emph{Missing Tranverse Energy:}  $\met> 400$ GeV.

 \item \emph{Invariant mass of Z-candidate:}  $85\ {\rm GeV}<m_{\ell^{+}\ell^{-}}<95\ {\rm GeV}$.

  \item \emph{Invariant mass of h-candidate:} $75\ {\rm GeV}<m_{bb}<150\ {\rm GeV}$ for two leading $b$-jets.

  \item \emph{Razor variables}:  Two razor variables are used in our analyses: 
\begin{align}
M_R &= \sqrt{(E_Z+E_h)^2 - (p_Z^z + p_h^z)^2},\\
M_T^R &= \sqrt{\frac{1}{2}\left[\slashed{E}_T(|\vec{p}_{ZT}|+|\vec{p}_{hT}|)
- \vec{\slashed{E}}_T\cdot(\vec{p}_{ZT}+\vec{p}_{hT})\right]}
\end{align}
 for $E_{Z,h}$ and $\vec{p}_{Z,h}$ being the reconstructed energy and momentum for $Z$ and $h$, respectively.
      Representative kinematic
  distributions of these variables for a signal benchmark point with $\mu=1$ TeV and $M_1=25$ GeV, as well as $tt$ and $tbW$ backgrounds are
  shown in Fig.~\ref{fig:razor_histos}, after the trigger and identification cuts.
  The distribution for the \emph{bbWW} background is not shown, since it is negligible in
  comparison to the others.   We can see that the distribution of $M_R$ is peaked
  around 1 TeV for the signal, which is consistent with what we would expect as
  it corresponds to the mass difference between the NLSPs and the LSP.  The $tt$
  background peaks around 500 GeV, while the distribution of $tbW$ background
  almost overlaps with that of the signal process.
  Similar behaviour is also seen in the $M_T^R$ distributions.  
  For this mass combination, requiring $M_R >$ 800
  GeV and $M_T^R >$ 400 GeV yields the greatest significance.

\end{enumerate}

\begin{table}[h]
  \centering
  \begin{tabular}{lrrrrrrr}
\toprule
{} &  $\sigma_\text{\emph{signal}}$ &  $\sigma_{t\overline{t}}$ &  $\sigma_\text{\emph{tbW}}$ &  $\sigma_\text{\emph{bbWW}}$ &  $\sigma_\text{\emph{BG (total)}}$ &   $S/B$ &  $S/\sqrt{B}$ \\
\midrule
Original            & 0.37 & 35,998 & 4,176 & 7.8     & 40,182 & 9.1$\times$ 10$^{-6}$ & 0.10 \\
Trigger             & 0.31 & 5,321  & 1,058 & 2.5     & 6,382  & 4.9$\times$ 10$^{-5}$ & 0.21 \\
\textsc{sfos} leptons        & 0.25 & 1,774  & 360   & 0.88    & 2,135  & 1.2$\times$ 10$^{-4}$ & 0.30 \\
2 $b$ jets          & 0.04 & 290    & 62    & 0.09    & 352    & 1.3$\times$ 10$^{-4}$ & 0.13 \\
$\slashed{E}_T >$ 400         & 0.03 & 5.3    & 6.8   & 0.007   & 12     & 0.003               & 0.49 \\
$m_{\ell\ell} \in$ [85, 95]  & 0.03 & 2.1    & 3.3   & 0.004   & 5.3    & 0.005               & 0.62 \\
$m_{bb}\in$ [75,150] & 0.02 & 0.59   & 0.30  & 8.2$\times$ 10$^{-4}$ & 0.90   & 0.02                & 1.3 \\
$M_{R} >$ 800       & 0.02 & 0.03   & 0.20  & 3.3$\times$ 10$^{-4}$ & 0.23   & 0.09                & 2.2 \\
$M_{T}^{R} >$ 400   & 0.02 & 0.008  & 0.18  & 1.9$\times$ 10$^{-4}$ & 0.19   & 0.10                & 2.4 \\
\bottomrule
\end{tabular}

  \caption{Representative cut flow table for a signal benchmark point with $\mu=1$ TeV,
    $M_1 = 25$ GeV at 100  TeV $pp$ collider, for a traditional cut-and-count analysis. All cross sections
    are given in unit of fb, and the units for the missing energy, invariant
    mass, and razor variable cuts are GeV. The significance, $S/\sqrt{B}$, is
    calculated for an integrated luminosity of 3 ab$^{-1}$.  }
\label{tab:cc_cutflowtable}
\end{table}

Table \ref{tab:cc_cutflowtable} shows the cut efficiencies for a representative
signal benchmark point with $\mu =$ 1 TeV and $M_1$ = 25 GeV.   The
cross-sections of the backgrounds are obtained from MadEvent after performing the MLM matching procedure. Their small size reflects
the $m_{\ell\ell}$ and $\slashed{E}_T$ cuts we imposed at the parton level. For all
the benchmark points, we require a minimum of 3 signal events left over after
cuts. We can see that, after applying our cuts, $tbW$ remains as the
dominant backgrounds, followed by $tt$.   The values of the razor variable cuts were chosen to
maximize the significance, $S/\sqrt{B}$ (calculated for an integrated luminosity
of 3000 fb$^{-1}$), shown in the last column.  

\subsection{Analysis using gradient boosted decision trees}\label{subsec:bdt}

For each signal mass combination and each background process, we preselected
events that passed the lepton trigger, contained two SFOS leptons and two
$b$-tagged jets, and calculated a number of kinematic variables for each event.   We then
placed these kinematic variables in an array, with each row corresponding to an
event, and the columns corresponding to the kinematic variables. A mixture of
low-level and high-level kinematic variables was shown to have the greatest
effectiveness. The kinematic variables chosen were $m_{\ell\ell}$, $m_{bb}$,
$M_R$, $M_T^R$, $\slashed{E}_T,$ the total hadronic transverse energy $H_T$,
and the transverse momenta of individual final state leptons and $b$-quarks:
$p_T^{\ell_1}$, $p_T^{\ell_2}$, $p_T^{b_1}$, and $p_T^{b_2}$. We then divided
the generated events into training and test sets, with the training set
comprising 75\% of signal events and 30\% of background events.  We used the
\texttt{scikit-learn} package  \cite{Pedregosa2011} to implement our analysis.
A boosted decision tree classifier was trained with 1000 weak learners and a
learning rate of 0.025.  

\begin{figure}[h]
\centering
\includegraphics[trim = {0 0.5cm 0 0},clip]{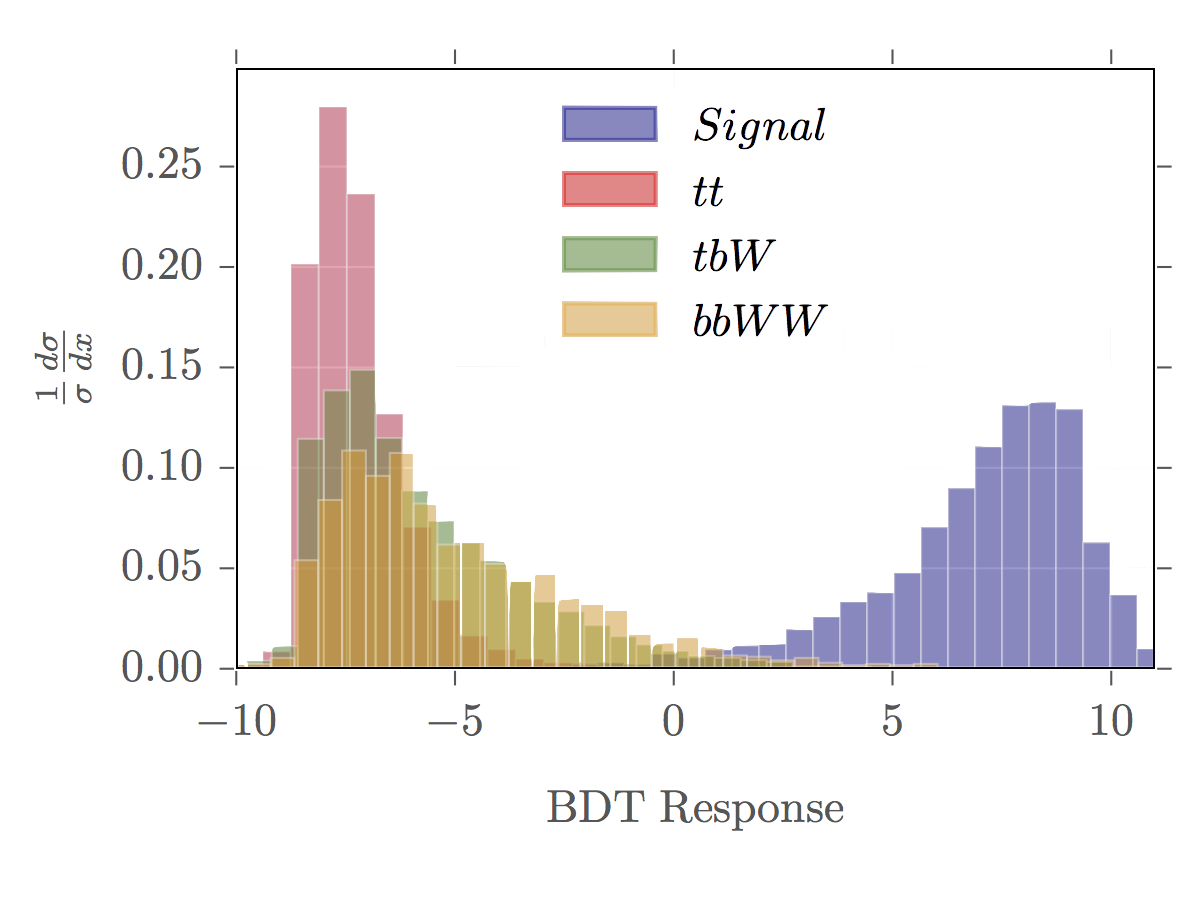}
\caption{Distribution of the decision function of the gradient boosted decision
tree classifier for   the representative signal benchmark point ($\mu = 1$ TeV, $M_1 = 25$ GeV) and backgrounds.}
\label{fig:bdt_response}
\end{figure}

After the classifier is trained, we use it to assign scores to individual
events. A more negative score indicates a more background-like event, and a
more positive score denotes a more signal-like event. After the scores have
been assigned to the events in the test sets, we can use this score just as we
would a regular kinematic variable, that is, we apply a cut that selects
events with a minimum of this score. The value of the cut is chosen to maximize
the significance for each signal benchmark point. The distribution of scores
for the representative benchmark point ($\mu$ = 1 TeV, $M_1$ = 25 GeV) is shown in Fig.~\ref{fig:bdt_response}. We observe that there is an appreciable
separation between the signal and background distributions.

\begin{table}[h]
  \centering
  \begin{tabular}{lrrrrrrr}
\toprule
{} &  $\sigma_\text{\emph{signal}}$ &  $\sigma_{t\overline{t}}$ &  $\sigma_\text{\emph{tbW}}$
   &  $\sigma_\text{\emph{bbWW}}$ &  $\sigma_\text{\emph{background (total)}}$ &   $S/B$ &  $S/\sqrt{B}$ \\
\midrule
Original           &               0.37 &          35,998 &           4,176 &              7.8 &             40,182 & 9.1e-06 &          0.10 \\
Preselection &               0.04 &             290 &              62 &             0.09 &                352 & 1.3e-04 &          0.13 \\
\textsc{bdt} response $> 5.1$      &               0.04 &            0.02 &            0.04 &          4.8e-04 &               0.06 &    0.63 &           8.4 \\
\bottomrule
\end{tabular}

  \caption{Representative cut flow table for the same benchmark point and
    integrated luminosity as in Table \ref{tab:cc_cutflowtable}, but using a BDT  analysis instead. The preselection is equivalent
    to the trigger and identification cuts listed in Table
    \ref{tab:cc_cutflowtable}. As before, all the cross sections are in
  fb.  }
\label{tab:bdt_cutflowtable}
\end{table}

Table \ref{tab:bdt_cutflowtable} shows the cut efficiencies for the same
representative benchmark point as in Table \ref{tab:cc_cutflowtable}, but this
time for a boosted decision tree analysis. We observe that the statistical
significance we can achieve goes up from 2.4 to 8.4, a roughly four-fold
increase.  

For certain values of the
cuts on kinematic variables or the score of the classifier, no events survived
for one or more of the background components. To guard against overly optimistic
estimates of the significance $S/\sqrt{B}$, we set a lower bound of 3 for the
number of Monte Carlo (MC) events after cuts for each background component, corresponding to
the 95\% confidence interval for a Poisson distribution.

\begin{figure}[h]
\centering
\input{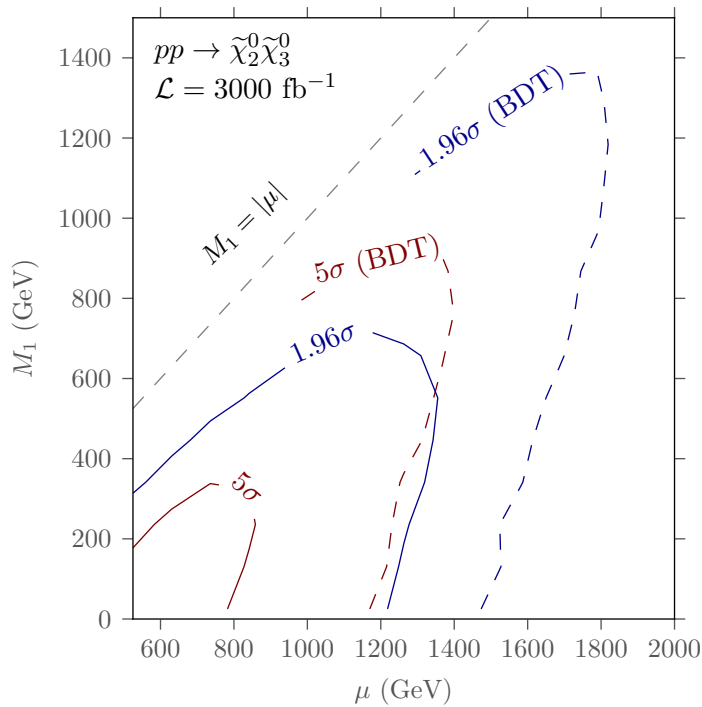}

\caption{ Discovery (red) and exclusion (blue) contours for the traditional
  cut-and-count analysis (solid) and boosted decision tree analysis (dashed),
  for 100 TeV $pp$ collider with an integrated luminosity of 3000 fb$^{-1}$.
}

\label{fig:contours}
\end{figure}

\section{Discovery and exclusion limits}
\label{sec:results}

Fig.~\ref{fig:contours} shows the expected reach  for 95\% C.L. exclusion and
5$\sigma$ discovery in the parameter space of $\mu$ vs.  $M_1$ for 100 TeV $pp$
collider with an integrated luminosity of 3000 fb$^{-1}$. The cut-and-count
strategy is able to discover Higgsinos up to 850 GeV  and exclude them up to
1.35 TeV.  The boosted decision
tree analysis improves this reach -- it is able to discover Higgsinos up to
1.4 TeV, and exclude them up to 1.8 TeV in favorable points in parameter space.
Similarly, Binos can be discovered up to 350 GeV and excluded up to 700 GeV
with the rectangular cut analysis, and the BDT analysis can discover them up to
about 900 GeV and exclude them up to 1.4 TeV. Since the background left after
cuts is relatively low, we do not include systematic errors in these estimates.

Comparing the BDT results with that of the cut-and-count bases analysis, the improvement is substantial, but perhaps not spectacular. However, it should
be kept in mind that this is probably a conservative estimate. Since we reserve
30\% of the background data for training purposes, and impose the condition that the
minimum number of MC events after cuts for the backgrounds must be 3, the true rate of background
rejection by the classifier is likely underestimated, and the expected
significance from the BDT analysis will improve. This extrapolation is
qualitatively supported by the high degree of separation between signal and
backgrounds seen in Fig.~\ref{fig:bdt_response}, and the fact that a larger
fraction of the signal events are preserved with the BDT approach compared to
the rectangular cut approach, as can be seen in Tables \ref{tab:cc_cutflowtable}
and \ref{tab:bdt_cutflowtable}.  

In regions with large mass differences between the LSP and NLSP, the reach in $\mu$ is
reduced.  This is because the intermediate SM Higgs boson will be highly boosted, and
thus decay into a highly collimated pair of $b$-jets. The extent to which this
affects our results can be estimated from Table \ref{tab:cc_cutflowtable}. We
see that requiring two $b$-tagged jets reduces our signal cross-section by
roughly 87\%. This is far more than the 28\% reduction that one might naively
expect as a result of applying the $b$-tagging efficiency of about 85\%
specified in our \texttt{Delphes} card.  
 
For massless Binos, the reach of the $Zh+\met$ channel in our study is less than the reach of the multi-lepton channel studied in Ref.~\cite{Gori:2014oua},
which can exclude Higgsinos with masses up to 2.8 TeV and discover them for masses 
up to 1.8 TeV.
However, for heavier Binos (larger $M_1$), the reach of the multi-lepton channel
is reduced, and vanishes completely for $M_1\gtrsim 600$ GeV (1000 GeV)
for discovery (exclusion). In this region, the reach of the $Zh+\met$ channel in our study exceeds that of the multi-lepton search for larger LSP masses: an increasement of about 300 GeV with BDT analysis.

\section{Conclusion}\label{sec:conclusion}

In this paper, we examined the discovery/exclusion potential of a 100 TeV \emph{pp} collider
for pair-produced heavy Higgsino NLSPs with a Bino LSP, via intermediate \emph{Z} and
\emph{h} bosons that decay to a pair of leptons and \emph{b} quarks respectively:
$\N_2\N_3\rightarrow Zh \N_1\N_1\rightarrow bb\ell\ell \met$.

We pursued two analysis strategies. One was the traditional cut-and-count methods, including razor variables
that are sensitive to the mass difference between the LSP and the NLSPs. The
other strategy was to use a boosted decision tree classifier trained with a
number of low- and high-level kinematic variables, including razor variables. We expected
that the machine learning approach would be able to more efficiently determine
the optimal decision boundary between signal and background events in feature
space than the traditional   cut-and-count method.

Overall, we find that the reach of our analysis strategy is a significant
improvement over that of the LHC. We found that the rectangular cut strategy has
the potential to discover Higgsino NLSPs up to a mass of 850 GeV, and exclude
them up to a mass of 1.35 TeV, for a Bino LSP mass around 350 GeV and 700 GeV
respectively. 
As expected, the boosted decision tree classifier
performs better, with the ability to discover Higgsino NLSPs up to 1.4 TeV and
exclude them up to a mass of 1.8 TeV, for a Bino mass of about 900 GeV and 1.4
TeV respectively.   In the
process, we highlighted the importance of generating enough MC  events
to estimate the huge backgrounds at a 100 TeV $pp$ collider.   

Additionally, we found that the reach for both strategies is considerably
reduced when the difference $ \mu-M_1$ is high, since it results in a highly
boosted $h$ that decays to a collimated pair of $b$-jets that will most likely be
identified as a single jet. This is not an insurmountable difficulty - there are
ways to deal with collimated jets, although they are beyond the scope of this
work. This issue does, however, highlight the necessity of improving isolation
performance at a 100 TeV $pp$ machine, where large mass hierarchies can result in
highly boosted/collimated decay products. For a review of currently used methods
to determine jet substructure, see \cite{Shelton:2013an}. In the future,
machine learning techniques might be profitably applied to this area as well -
for a review of developments along this line, see \cite{Schwartzman:2016jqu}. 

The collimation of the $b$ jets, along with the fact that we simulate
detector effects using Delphes, result in a relatively lower reach in the high
$\mu$, low $M_1$ region than the multilepton analysis in~\cite{Gori:2014oua},
which is able to exclude Higgsinos up to a mass of 2.8 TeV for massless Binos,
or discover them up to a mass of 1.8 TeV. However, the reach of our analysis
is higher in the region where the difference between $\mu$ and $M_1$
is smaller. This is consistent with our usage of the razor variable $M_R$,
which is sensitive to the mass difference between the parent Higgsinos and the
daughter Bino.

A 100 TeV \emph{pp} collider represents an excellent opportunity to discover
physics beyond the Standard Model. The extremely high energies and luminosities
involved will present new challenges for particle physicists, and it is likely
that machine learning will play an important part in facing them. 
  
\acknowledgments

We would like to thank Matt Leone and Ken Johns for helpful discussions. The
research activities of AP and SS were supported in part by the Department of
Energy under Grant DE-FG02-13ER41976/de-sc0009913. An allocation of computer
time from the UA Research Computing High Performance Computing (HPC) and High
Throughput Computing (HTC) at the University of Arizona is gratefully
acknowledged. We also thank KITP for its hospitality when this draft is completed.
This research was supported in part by the National Science Foundation under Grant No. NSF PHY-1748958.

\bibliography{references}

\end{document}